\RequirePackage{fix-cm}
\documentclass[smallextended]{svjour3}       
\smartqed
\usepackage{graphicx}
\usepackage{latexsym}
\usepackage{epsfig,amsmath,amssymb,amsfonts,bm,lineno}
\usepackage{array}
\usepackage{parskip}
\usepackage{color}

\usepackage{latexsym}

\newcommand{\ds}{\displaystyle}
\newcommand{\beq}{\begin{equation}}
\newcommand{\eeq}{\end{equation}}
\newcommand{\beqq}{\begin{eqnarray*}}
\newcommand{\eeqq}{\end{eqnarray*}}

\begin{document}

\title{Stochastic model of endosomal escape of Influenza virus}
\author{
Thibault Lagache \and
Christian Sieben \and
Tim Meyer \and
Andreas Herrmann \and 
David Holcman}
\institute{T. Lagache \at Applied Mathematics and Computational Biology, IBENS, Ecole Normale Sup\'erieure, 46 rue d'Ulm 75005 PARIS, France \\   \email{thibault.lagache@pasteur.fr} \\   \emph{Present address:} BioImage Analysis Unit, CNRS UMR 3671, Institut Pasteur, France. 
\and C. Sieben \at Department of Biology, Molecular Biophysics, IRI Life Sciences, Humboldt-Universit\"at zu Berlin, Germany
\and T. Meyer \at Institute of Chemistry and Biochemistry, Free University Berlin, Germany
\and A. Herrmann \at Department of Biology, Molecular Biophysics, IRI Life Sciences, Humboldt-Universit\"at zu Berlin, Germany
\and D. Holcman \at  Applied Mathematics and Computational Biology, IBENS, Ecole Normale Sup\'erieure, 46 rue d'Ulm 75005 PARIS, France and Newton Institute,DAMTP Cambridge Cb3 0DS and Mathematical Institute, University of Oxford, Andrew Wiles Building, Woodstock Rd, Oxford OX2 6GG, United Kingdom \\   \email{david.holcman@ens.fr}}

\date{Received: XX/ Accepted: XX}
\maketitle
\begin{abstract}
\end{abstract}

\section{Abstract}
Influenza viruses enter a cell via endocytosis after binding to the surface. During the endosomal journey, acidification triggers a conformational change of the virus spike protein hemagglutinin (HA) that results in escape of the viral genome from the endosome to the cytoplasm. A quantitative understanding of the processes involved in HA mediated fusion with the endosome is still missing. We develop here a stochastic model to estimate the change of conformation of HAs inside the endosome nanodomain. Using a Markov-jump process to model the arrival of protons to HA binding sites, we compute the kinetics of their accumulation and the mean first time for HAs to be activated. This analysis reveals that HA proton binding sites possess a high chemical barrier, ensuring a stability of the spike protein at sub-acidic pH. Finally, we predict that activating more than $3$ adjacent HAs is necessary to prevent a premature fusion.
\section{Introduction}
The first step of infection by influenza starts when viral particles enter the cell by a process called endocytosis at the host cell surface, where they are captured in spherical endosome. The second step is the transport of the virus, trapped inside the endosome. During the third and critical step, the viral genome, encoding ribonucleoproteins  (vRNPs) has to escape from the endosomal compartment, so that later on, it can translocate into the nucleus \cite{Mercer:2010fk} (Figure \ref{figure1}-A). Fusion between the endosome and influenza virus  is mediated by a low-pH conformational change of the viral envelope glycoprotein hemagglutinin (HA) (figure \ref{figure1}-A). The goal of this article is to present a new model based on endosome acidification and conformational change of the HA to predict the exact timing for initiating fusion between the virus and the endosome membrane, and thus to release the viral genome.

The model accounts for important detailed properties of the glycoprotein HA composed by two linked subunits HA1 and  HA2, the latter anchoring HA to the viral envelope. Indeed, at neutral pH, HA is not active (in a non-fusogenic state), but as the pH decreases due to proton entry into the endosome, a partial dissociation of the HA1 subunit results in a spring-loaded conformational change of HA2 into an active  (fusogenic) state \cite{Huang:2003qf}. Consequently, the  residence time of influenza virus genome within the endosome before fusion depends on the kinetics of endosome acidification.  Yet, the absence of direct in vivo measurements of these parameters makes the endosomal  step of virus infection difficult to analyze. To estimate the pH-driven fusion of influenza viruses in an endosome, we develop a model that accounts for the main kinetics parameters of the fusion process.

We start by developing a kinetics model for endosome acidification, that we calibrate using experimental data. The model depends on the following parameters: buffering  capacity of the endosomal lumen, membrane leakage and proton pumping rate, all together controlling the number of free protons inside the endosome. Because the proton binding event to HA is discrete, we model it here using a classical Markov jump process \cite{Schuss:2010}. Using an asymptotic expansion of the solution for the mean arrival time equation for the number of protons to a certain threshold, we obtain an analytical expression for the kinetics of HA conformational change at fix pH values. The model is then calibrated using the kinetics of the HA conformational change \cite{Krumbiegel:1994bh}. Finally, by combining the two models for kinetics of endosomal acidification and HA conformational change, we can estimate the number of activated HAs inside an endosome. We predict that at least three adjacent activated HAs are necessary to trigger membrane fusion \cite{Danieli:1996ve,Ivanovic:2013fk}, a cooperativity process that should prevent premature fusion. We confirm some of the predictions using co-labeling viruses and endosomal markers experimental data, showing that intracellular fusion of viruses mainly occur in maturing endosomes.


\section{Kinetics model of endosomal acidification}
The model of endosomal acidification follows the free number of protons $P_e(t)$ at time $t$ in the endosomal compartment. The protons enter with an entry rate $\lambda(t) S$ through the V-ATPase proton pumps ($S$ is the endosomal surface and the rate $\lambda(t)$ is associated with to the proton pumps activity) and can escape with a leaking rate $L_{ext}(t)$, but can also bind to endosomal buffers. The proton pump rate $\lambda(t)$ is mainly determined by the membrane potential $\Psi(t)$ (Figure 11 \cite{Grabe:2000vn}), which in turn depends on the endosomal concentrations of several cations ($H^{+}, K^{+}, Na^{+}\ldots$) and ($Cl^{-}\ldots$). The ionic concentration inside endosome is tightly regulated by channels, exchangers and leak and in particular, by raising the interior-positive membrane potential, Na-K ATPase exchangers have been proposed to limit the acidification of early compared to late endosomes \cite{Fuchs:1989ly}.

\subsection{Mass action law for free protons}
To derive the time-dependent equations for the free protons, we use the balance of fluxes. The fast equilibrium between fluxes determines the effective number of protons $P_e(t) dt$ entering the endosome which follows the first order kinetics
\beq
\frac{d P_e}{dt} =\left(\lambda(t) S-L_{ext}(t)\right).\label{deltaP}
\eeq
Protons are rapidly bound to endosomal buffers. We model the buffer activity using an ensemble of chemical reactions $\left(P_eB_i/B_i\right)_{1\leq i \leq n}$ \cite{Grabe:2001fk}:
\beq
P_e+B_1 \overset{k_1}{\underset {k_{1}^{(-1)}}{\rightleftharpoons}} P_eB_1 \hbox{  },P_e+B_2 \overset{k_2}{\underset {k_{2}^{(-1)}}{\rightleftharpoons}} P_eB_2
\ldots \hbox{  },P_e+B_n \overset{k_n}{\underset {k_{n}^{(-1)}}{\rightleftharpoons}} P_eB_n, \label{acidbase}
\eeq
where $\left(k_i,k_i^{(-1)}\right)_{1 \leq i \leq n}$ are the rate constants  $\left(P_eB_i/B_i\right)_{1\leq i \leq n}$. Thus the kinetics equations for the number of free protons $P_e(t)$ inside an endosome is
\beqq
\frac{dP_e(t)}{dt} &=& \Delta P_e(t)+\sum_{i=1}^{n}\left(k_{i}^{(-1)} P_eB_i(t)-\frac{k_i}{N_A V_e} P_e(t) B_i(t)\right)\\
&=& \left(\lambda(t) S-L_{ext}(t)\right)+\sum_{i=1}^{n}\left(k_{i}^{(-1)} P_eB_i(t)-\frac{k_i}{N_A V_e}  P_e(t) B_i(t)\right), \label{differential0}
\eeqq
where $P_eB_i(t)$ and $B_i(t)$ are the number of weak acids and bases inside the endosome at time $t$. We assume that the membrane potential $\Psi(t)$ reaches rapidly its steady state value $\Psi(\infty)$ compared to the acidification kinetics \cite{Grabe:2000vn}, we thus approximate the pumping rate $\lambda(t) S$ as
\beq
\lambda(t)S\approx \lambda S.\label{lambda}
\eeq
where the parameter $\lambda$ is related to the membrane potential $\Psi(\infty)$. In addition, the protons leak $L_{ext}(t)$ is proportional to the endosomal concentration and the endosomal surface \cite{Grabe:2001fk}
\beq
L_{ext}(t)=LS\frac{P_e(t)}{N_AV_e},\label{Lextru}
\eeq
where $L$ is a permeability constant, $N_A$ is the Avogadro constant and $V_e$ is the volume of the endosome. Consequently, using approximations \ref{lambda}-\ref{Lextru} in equation \ref{differential0}, we obtain the general dynamics of free protons:
\beqq
\frac{dP_e(t)}{dt} &=&\left(\lambda - L\frac{P_e(t)}{N_AV_e}\right)S+\sum_{i=1}^{n}\left(k_{i}^{(-1)} P_eB_i(t)-\frac{k_i}{N_A V_e} P_e(t) B_i(t),\right) \label{differential2}.
\eeqq
\subsection{Dynamics of the pH and the steady state limit}
When the protons enter the endosome, they equilibrate with buffers, a process much faster compared to acidification: after a fraction of $\Delta P_e(t)$ protons have entered the endosome, they instantaneously bind to bases, leading to a jump $-\Delta B_i(t)$ on each base
\beq
\Delta P_e(t) \approx -\sum_{i=1}^{n}\Delta B_i(t).\label{dPe}
\eeq
To estimate the associated pH change $d\text{pH}$ with the entry $\Delta P_e(t)$ of protons and the infinitesimal changes $-dB_i(t)$ of the number of bases, we use equation \ref{acidbase}) at equilibrium and time $t$:
\beq
k_i^{(-1)} P_eB_i(t)=k_i \frac{P_e(t)B_i(t)}{N_A V_e} \hbox{, for all }1 \leq i \leq n.
\eeq
Thus,
\beq
\frac{P_e(t)}{N_A V_e} = K_i \frac{C_i-B_i(t)}{B_i(t)} \hbox{, for all } 1 \leq i \leq n, \label{Ki}
\eeq
where $K_i=\frac{k_i^{(-1)}}{k_i}$ and $C_i=P_eB_i(0)+ B_i(0)$ are constant. Consequently,
\beq
\text{pH}(t) = pK_i+\frac{1}{\log(10)} \log\left(\frac{B_i(t)}{\left(C_i-B_i(t)\right)}\right) \label{pH}
\eeq
where $pK_i=-\log(K_i)/\log(10)$. By differentiating equation \ref{pH} with respect to $B_i$, the infinitesimal variation $dB_i$ of base $i$ is related to $d\text{pH}$ of the endosomal pH by
\beq
d\text{pH}=\left(\frac{1}{\log(10)}\frac{C_i}{B_i(t)\left(C_i-B_i(t)\right)}\right)dB_i.\label{d1}
\eeq
Using equation \ref{Ki}, we get for equation \ref{d1}
\beq
d\text{pH}=\left(\frac{1}{\log(10)}\frac{\left(N_AV_eK_i+P_e(t)\right)^2}{P_e(t)C_iLN_AV_e}\right)dB_i,\label{d2}
\eeq
leading to
\beq
dB_i=N_A V_e \beta_i\left(P_e(t)\right)d\text{pH},\label{d3}
\eeq
where
\beq
\beta_i\left(P_e(t)\right)=\log(10)C_i \frac{P_e(t) K_i}{\left(P_e(t)+K_i N_AV_e\right)^2}
\eeq
is the buffering capacity of the weak acid-base $\left(P_eB_i, B_i\right)$. Finally, using equations \ref{dPe} and \ref{d3} we find that the variation $\Delta P_e(t)$ of protons is related to an infinitesimal change $\Delta \text{pH}$ of the endosomal pH through
\beq
\Delta P_e(t)=-\sum_{i=1}^{n}\Delta B_i=-N_A V_e \left(\sum_{i=1}^{n}\beta_i\left(P_e(t)\right)\right) \Delta\text{pH}=-N_A V_e \beta_e^0 \left(P_e(t)\right) \Delta\text{pH},\label{dPe1}
\eeq
where $\beta_e^0\left(P_e(t)\right)=\sum_{i=1}^{n}\beta_i\left(P_e(t)\right)$ is the total buffering capacity of the endosome, which is approximately constant  $\beta_e^0\left(P_e(t)\right) \approx \beta_e^0=40 mM/pH$\cite{Van-Dyke:1994bh}. \\
Finally, using the proton extrusion and pumping rates (equation \ref{lambda} and \ref{Lextru}), we obtain the kinetics equation
\beq
\left(\lambda - L\frac{P_e(t)}{N_AV_e}\right)S dt =-N_A V_e \beta_e d\text{pH}.\label{d5}
\eeq
With
\beq
\frac{d\text{pH}(t)}{dt}=-\frac{1}{\log(10)P_e(t)} \frac{dP_e(t)}{dt},
\eeq
we obtain that the kinetics equation for free protons accumulation in an endosome is
\beq
\frac{dP_e(t)}{dt}=\left(\lambda - L\frac{P_e(t)}{N_AV_e}\right) \frac{S\log(10)P_e(t)}{N_A V_e \beta_e}\label{acidif}.
\eeq
When the proton leakage is counterbalanced by the pumps, after a time long enough, the pH reaches an asymptotic value $pH_{\infty}$, where the endosome cannot be further acidified given by
\beq
P_e(\infty)=N_A V_e 10^{-\text{pH}_{\infty}},
\eeq
Consequently, the rate $\lambda$ is linked to $pH_{\infty}$ by
\beq
\lambda=L 10^{-\text{pH}_{\infty}},
\eeq
and equation \ref{acidif} can be rewritten as
\beq
\frac{dP_e(t)}{dt}=\left(10^{-\text{pH}_{\infty}} - \frac{P_e(t)}{N_AV_e}\right)\frac{LS\log(10)P_e(t)}{N_A V_e \beta_e}\label{acidif2}.
\eeq
To conclude, we obtain here a kinetics equation for the endosome acidification as a function of endosome parameters and permeability. However, equation \ref{acidif2} is not sufficient to account for the different stages of the endosomal maturation. Indeed, the final  $pH_{\infty}$ \cite{Mercer:2010fk} and the permeability $L$ were reported to decrease with the endosomal maturation \cite{Fuchs:1989ly} and are thus time dependent.

\subsection{Modeling pH change and acidification of an endosome}
We now relate the acidification dynamics to the change of two proteins that can be followed experimentally. Indeed, the transition from a first stage endosome called early endosome (EE) to a second stage endosome called late endosome (LE) is quantified by a gradual exchange of a protein called Rab5 by another one associated to the late phase called protein Rab7 \cite{Rink:2005kl}. Kinetics of Rab exchange have been experimentally measured and we approximate here the kinetics of the ratio Rab5/Rab7 obtained from data (Figure 4-C \cite{Rink:2005kl}) by a sigmoidal function
\beq
\frac{\text{Rab7(t)}}{\text{Rab5(t)}+\text{Rab7(t)}}=\frac{1}{1+e^{-(t-t_{1/2})/\tau_c}},\label{RabC}
\eeq
where $t_{1/2}$ is the half-maturation time and $\tau_c$ is the time scale of Rab conversion. We then approximate the transition rate from early to late endosome with the Rab conversion rate and we consider that the steady-state pH$_{\infty}(t)$ relative to the amount of Rab7, that gradually replaces Rab5 during endosomal maturation is given by
\beq
\text{pH}_{\infty}(t) =\text{pH}_{\infty}^{\text{early}}+\left(\text{pH}_{\infty}^{\text{late}}-\text{pH}_{\infty}^{\text{early}}\right) \frac{\text{Rab7(t)}}{\text{Rab5(t)}+\text{Rab7(t)}}.\label{pHRab}
\eeq
Similarly, the permeability constant follows the equation
\beq
\text{L}(t) =L^{\text{early}}+\left(L^{\text{late}}-L^{\text{early}}\right) \frac{\text{Rab7(t)}}{\text{Rab5(t)}+\text{Rab7(t)}}.\label{KRab}
\eeq
\subsection{Calibrating the acidification model by live cell imaging} \label{s:acidification}
We shall now calibrate the acidification equation to experimental parameters. First, by fitting equation \ref{RabC} to the experimental data (Figure4-C of \cite{Rink:2005kl}) where the lag time between initiation and termination of the Rab5/Rab7 replacement is estimated to $\approx 10$ min., leading to a time constant for $\tau_c\approx$ 100 s.

We use data from endosomal acidification in MDCK cells where the pH inside endosomes decreases very quickly within the first 10-15 min (Figure \ref{figure1}-B) to reach a steady state pH around 5.5 after 20 min, in agreement with previous studies \cite{Zaraket:2013aa}. The steady state pH is pH$_{\infty}^{\text{early}}=6.0$ and pH$_{\infty}^{\text{late}}=5.5$ for early and late endosomes, respectively \cite{Bayer:1998cr},  thus we calibrated the permeability constant $L$ and Rab conversion kinetics by solving numerically equation \ref{acidif2} and fitting the experimental acidification curve (Figure \ref{figure1}-C). We found that the permeabilities of early and late endosomes are $L^{\text{early}}\approx 3.5\hbox{ } 10^{-3} N_A \text{cm s}^{-1}$ and $L^{\text{late}}\approx 0.1\hbox{ }L^{\text{early}} = 3.5\hbox{ } 10^{-4} N_A\text{cm s}^{-1}$, respectively and the half-maturation time is $t_{1/2}=10$ min.
\subsection{Accounting for proton influx inside viral core and buffering} \label{protonBIG}
The last step of the kinetics model of protons include the binding and unbinding with various viral components providing buffer capacity. Indeed, the buffering capacity of the viral proteins and the influx of protons through M2-channels inside the viral core (Figure\ref{figure1}-A), the presence of viruses inside endosomes changes the overall buffering capacity of the endosome itself and can perturb the overall acidification kinetics.
 To compute the influx in each virus through M2 channels, we use first order transport kinetics \cite{Leiding:2010uq}, summarized by the chemical reaction
\begin{linenomath}
\beq
P_e+M2 \overset{k_e}{\underset {k_e^{-1}}{\rightleftharpoons}} M2-P  \overset{k_v^{-1}}{\underset {k_v}{\rightleftharpoons}} P_v+M2. \label{chem}
\eeq
\end{linenomath}
When a proton $P_e$ binds a free $M2$ protein channel with a binding  $k_e$ and unbinding  $k^{-1}_e$ rates, it is transported inside the virus core with an inward rate $k^{-1}_v$, while the forward rate is $k_v$. Thus the inward flux in a single virus can be computed from equation \ref{chem} \cite{Leiding:2010uq}
\beq
{\ds j_{M2}(P_e,P_v)=n_{M2} \frac{k^{-1}_e-\frac{k_eP_e\alpha(P_e,P_v)}{N_AV_e}}{1+\alpha(P_e,P_v)}}, \label{fluxM2}
\eeq
where $n_{M2}$ is the number of M2 channels per viral particle, $P_v$ is the number of free protons inside the viral core and
\beq
{\ds \alpha(P_e,P_v)=\frac{k_e^{-1}+k_v^{-1}}{k_e\left(\frac{P_e}{N_A V_e}+\frac{k_v^{-1}P_v}{k_e^{-1}N_A V_v}\right)}},
\eeq
We extracted the buffer capacity of a virus and accounted for the viral genome, the internal viral proteins and unspecific buffers that can be reached  through the M2 channels  \cite{Leiding:2010uq}. Most abundant internal proteins are the M1 ($3,000$ copies per virus) and the nucleoproteins (NP, $330$ copies per virus) \cite{LambandKrug} (Figure \ref{figure1}-A). Proton binding sites of viral proteins are the ionogenic groups in their amino acid side chains \cite{stoyanov}, and the main ionogenic buffers in the endosome pH range are the aspartic acid (Asp, pKa=3.9), the glutamic acid (Glu, pKa=4.32) and the histidine (His, pKa=6.04) \cite{stoyanov}.  Closely related binding sites can have strong influence on each other due to electrostatic interactions.

In addition, the 3-dimensional protein folding can hinder the accessibility of some residues to the solvent and protons. Consequently, calculations based on the three-dimensional structure of the protein are necessary to determine accurately its buffering capacity with respect to pH. Using the spatial organization (crystal structure) of NP proteins \cite{Ng:2008fk}, we computed the pKa values of all titratable residues in the proteins with electrostatic energy calculations using the software Karlsberg+ \cite{PROT:PROT21820}. We then determined the mean number of protonated residues $n_{P}^{NP}(pH)$ of NP proteins (see Material and Methods) and we found that $n_{P}^{NP}(pH)$ increases almost linearly with pH:
\beq
n_{P}^{NP}(pH)\approx n_{P}^{NP}(pH=7)+9 \left(7-pH\right),
\eeq
indicating that the buffering capacity of NPs is approximatively constant between pH 7 and 5 (equation \ref{d3})
\beq
\beta_{v}^{NP}\approx 9 \frac{330}{N_A V_v} = \frac{3000}{N_A V_v}
\eeq
where $V_v=\frac{4}{3}\pi r_v^3$ is the volume of the viral internal lumen, for a spherical viral particle with radius $r_v= 60$ nm \cite{Lamb:1983fk}. The structure of the matrix M1 protein is unknown and consequently, we use the cumulative contributions of Asp, Glu and His residues to estimate the number of M1 proton binding sites. We thus estimate the fraction $P_{i}(pH)$ of occupied residues for a fixed pH using the equilibrium constant $pKa_i$ for any residue $i$ (Asp, Glu or His) to be
\beq
P_i (pH)=\left(10^{pH-pKa_i}+1\right)^{-1}.
\eeq
The mean number $n_{P}^{M1}(pH)$ of protonated site is then given by
\beq \label{NP-M1}
n_{P}^{M1}(pH)=n^{M1}_{Asp} \left(10^{pH-3.9}+1\right)^{-1}+n^{M1}_{Glu}\left(10^{pH-4.32}+1\right)^{-1}+n^{M1}_{His}\left(10^{pH-6.04}+1\right)^{-1}.
\eeq
where the number of residue for each group is $n^{M1}_{Asp}=12$, $n^{M1}_{Glu}=12$ and $n^{M1}_{His}=5$. Using equation \ref{NP-M1}, we plotted $n_{P}^{M1}(pH)$as function of the pH and observed that $n_{P}^{M1}(pH)$ is almost a linear function
\beq
n_{P}^{M1}(pH)\approx n_{P}^{NP}(pH=7)+3.5 \left(7-pH\right),
\eeq
and obtain that
\beq
\beta_{v}^{M1}\approx 3.5 \frac{3000}{N_A V_v} = \frac{10500}{N_A V_v}.
\eeq
Additionally to internal M1s and NPs proteins, protons entering the viral core through M2 channels can also bind to viral nucleic acids and in particular to basic groups in the guanine, adenine and cytosine nucleotides \cite{stoyanov}. In particular, the buffering capacity $\beta^{RNA}$ of oligonucleotides in solution, for a concentration $c_{monomers}$ of monomers, has been estimated to be $\beta^{RNA} \approx 0.1\hbox{ } c_{monomers}$ in the pH range 5-7 (Figure 3-D in \cite{stoyanov}). Consequently the buffering capacity $\beta_v^{RNA}$ of the $\approx 12000$ viral nucleotides \cite{Hutchinson:2010ja} is approximatively equal to
\beq
\beta_v^{RNA}\approx 0.1 \frac{12000}{N_A V_v} = \frac{1200}{N_A V_v}.
\eeq
Finally, the viral core lumen should also contain other unspecific buffers such as cytoplasmic buffers enclosed during the viral assembly, leading to an unspecific buffering capacity $\beta^{0}_{v}(pH)$ inside the viral lumen that has to be added to the buffering capacities  $\beta_{v}^{NP}$ and  $\beta_{v}^{M1}$ of internal proteins. Due to possible ionic exchange between viral and endosomal lumens, we approximate $\beta^{0}_{v}(pH)$ with the endosomal buffering capacity $\beta_e^0$, which is independent of the pH  and has been estimated to be \cite{Van-Dyke:1994bh}
\beq
\beta_e^0 \approx 40 mM/\text{pH}.
\eeq

In summary, the proton buffering capacity inside viral cores is equal to
\beq
\beta_i = \beta_v^{0}+\beta_v^{M1}+\beta_v^{NP}+\beta_v^{RNA},
\eeq
and similarly to the flux equation \ref{acidif2}, the number of free protons $P_v(t)$ contained in viral core at time $t$, which determines the influx of protons through M2 channels (equation \ref{fluxM2}), follows the kinetics equation
\begin{linenomath}
\beq
\frac{dP_v(t)}{dt} =\frac{\log(10)}{N_A V_v \left(\beta_v^{0}+\beta_v^{M1}+\beta_v^{NP}+\beta_v^{RNA}\right)}P_v(t)j_{M2}\left(P_e(t),P_v(t)\right)\label{acidif_virus}.
\eeq
\end{linenomath}
By solving numerically equation \ref{acidif_virus}, we estimate that $\approx 60,000$ protons enter the viral core during endosomal maturation. Using endosomal acidification kinetics equation \ref{acidif2}, we estimate that more than $20,000,000$ protons bind to endosomal buffers during endosome acidification. Thus, the buffering capacity of a single virus should not influence the endosomal acidification. However the number of protons that bind to endosomal buffers drastically decreases with the endosomal size (e.g. $\approx 175,000$ for $r_e = 100$ nm instead of $r_e = 500$ nm). In addition viral particles may accumulate during the endosomal journey \cite{Matlin:1982fk}. Thus, for multiplicity of infection (MOI) and viral accumulation in endosomes, the viral buffering capacity may significantly affect the acidification kinetics of small and intermediate size endosomes.

\section{Markov jump model of HA conformational change}
Although the number of protons entering in the endosome is quite huge, as discussed in the previous section, the actual number of free protons defining endosomal pH is low ($\sim 300$ at pH 6 in an endosome with a radius of $500$ nm). In addition, there are few proton binding sites on a single HA that trigger its conformational change \cite{Huang:2002dq}, which is the event we shall monitor. This change of scale between many entering protons and the few free protons and HA binding sites requires a different description than the previous continuous model.
To compute the mean time for HA conformation to change as the pH drop, we shall first extract the forward and backward proton binding rates. For that purpose, we convert the HA conformational change kinetics, obtained from experimental data at various pH \cite{Krumbiegel:1994bh} into rate constants. 


At temperature $T=300K$, when the pH decreases from $7$ to $4$, the number of bound protons bound to HA1 increases approximatively from $123$ to $132$ (Figure 3 in \cite{Huang:2002dq}), suggesting that the number of available number of binding site is $n_s=9$ at acidic pH. The influenza virus carries $n_{HA}\approx 400$ HA trimers \cite{Ivanovic:2013fk} (Figure \ref{figure2}-A) and thus there are exactly $n_{HA} n_s$ sites that can competitively bind protons. The goal of this section is to compute the mean time that a threshold $n_T$ of bound protons to HA1 is reached, which is a model of fusogenic state, where the protein can engage into the generation of a fusion pore with the endosomal membrane.
\subsection{Modeling HA conformational change }
To analyse the conformational changes of a single HA trimer, we follow the occupied sites $X(t,c)$ at time $t$, for a fix proton concentration $c$. During time $t$ and $t+\Delta t$, the number of specific bound sites can either increase with a probability $r(X,c)\Delta t$, when a proton arrives to a free site or decreases with probability $l(X,c)\Delta t$ when a proton unbinds or remains unchanged with probability $1-l(X,c)\Delta t-r(X,c)\Delta t$ (Figure \ref{figure2}-A).  

We estimate hereafter the rates $l(X,c)$ and $r(X,c)$ and the critical threshold $n_T$. The forward rate  depends on the proton concentration $c$ and the number of free sites $n_s-X$ of the HA trimer, thus  
\begin{linenomath}
\beq \label{eqr}
r(X,c)=K c (n_s-X),
\eeq
\end{linenomath}
where $K$ is the forward binding rate of a proton to a binding site. 

To determine the proton unbinding rate $l(X,c)$, we use the values available for the HA1 protonation \cite{Huang:2002dq}. We approximate the number of bound protons $\tilde{X}_0(c) $ with respect to the proton concentration $c$ by a linear function (Figure3 \cite{Huang:2002dq}) 
\beq
\tilde{X}_0(c)=\tilde{X}_0\left(10^{-7} mol.L^{-1}\right)+X_0(c)=\tilde{X}_0\left(10^{-7} mol.L^{-1}\right)+\left(\frac73+\frac{\log(c)}{3\log 10 }\right)n_s,
\eeq
where $\tilde{X}_0\left(10^{-7} mol.L^{-1}\right)$ is the mean number of bound protons at pH=7 and
\beq
X_0(c)=\left(\frac73+\frac{\log(c)}{3\log 10 }\right)n_s \label{X0}
\eeq
is the mean number of HA1 sites that are additionally protonated for a proton concentration $c>10^{-7}$mol. L$^{-1}$. Because the unbinding rate does not depend on the proton concentration $l(X,c)=l(X)$ and we obtain for the equilibrium ratio $\frac{l(X,c)}{r(X,c)}=\frac{l(X)}{K c (n_s-X)}$. Using at equilibrium the concentration $c(X)=\ds{10^{{\frac{3X}{n_s}-7}}}$ for which $X_0(c(X))=X$, the mass-action law leads to $ \frac{l(X_0(c),c)}{r(X_0(c),c)}=1$  or equivalently $\frac{l(X)}{K c(X) (n_s-X)}=1$. Finally, we get
\beq
l(X)=K (n_s-X) \ds{10^{{\frac{3X}{n_s}-7}}}.
\eeq
In summary, the binding and unbinding rates $r$ and $l$ are given by 
\beq
r(X,c)=K c (n_s-X), \hbox{ and } l(X,c)=l(X)=K (n_s-X) \ds{10^{{\frac{3X}{n_s}-7}}}.\label{rl}
\eeq
\subsection{Rate of HA conformational change}
To compute the mean time that exactly $n_T$ protons are bound to a single HA we use a Markov jump process description. The Master equation is derived by evaluating during time $t$ and $t+\Delta t$, the variation in the number of bound sites $X(t,c)$ among the $n_s=9$ HA1 proton binding sites. The scaled variable is 
\beq
x(t,c)=\epsilon X(t,c),
\eeq 
where $\epsilon=1/n_s$ and using the difference $\Delta x=x(t+\Delta t,c)-x(t,c)$, the transition probabilities are 
\begin{eqnarray*}
Prob\{\Delta x=\epsilon|x(t,c)=x\}=r(x,c)\Delta t,\\
Prob\{\Delta x=-\epsilon|x(t,c)=x\}=l(x,c)\Delta t,\\
Prob\{\Delta x=0|x(t,c)=x\}=\left(1-r(x,c)-l(x,c)\right)\Delta t.
\end{eqnarray*}
At a fixed proton concentration, the probability $p(y,t|x,c)$ that the number of protonated sites is equal to $y$ at time $t$, that is $x(t,c)=y$), when there are initially $x$ bound sites ($x(t=0,c)=x$), is solution of the backward-Kolmogorov equation \cite{Matkowsky:1984,KNESSL:1984ij,KNESSL:1985bs,Schuss:2010,Lagache:2012vn} 
\begin{eqnarray}
p(y,t|x,c)&=&p(y,t-\Delta t|x+\epsilon,c)r(x,c)\Delta t
+p(y,t-\Delta t|x-\epsilon,c)l(x,c)\Delta t \nonumber \\&+&
p(y,t-\Delta t|x,c)(1-r(x,c)\Delta t-l(x,c)\Delta t),
\end{eqnarray}
which has the classical Kramers-Moyal expansion \cite{risken}
\begin{eqnarray}
\frac{ \partial p}{\partial t} (y,t|x,c) &=&L_x
p=r(x,c)\sum_{n=1}^{\infty}\frac{\epsilon^n}{n!}\left(\partial_x\right)^n
p(y,t|x,c) \nonumber \\ &+&
l(x,c)\sum_{n=1}^{\infty}\frac{\left(-\epsilon\right)^n}{n!}\left(\partial_x\right)^n
p(y,t|x,c).\label{BKME}
\end{eqnarray}
{The mean first time $\tau(x,c)$ that the process $X(t,c)$ reaches the threshold $x_T=n_T/n_s$ models the HA1 subunit filled with $n_s$ protons. It is precisely the mean of the first passage time for the bound protons $x(t,c)$ to reach the level $x_T$. The mean first passage time $  \tau(x,c)  =E[\tau|x(t=0,c)=x]$, satisfies \cite{Schuss:2010}:
\beq
L_x \tau(x,c)=-1 \hbox{ for } x \hbox{ in }[0,x_T],
\eeq
with the boundary conditions 
\beq
\tau(x,c)=0 \hbox{ for } x=x_T \hbox{ and }\frac{\partial \tau(x,c)}{\partial x}= 0 \hbox{ for }x=0.
\eeq
{For $\epsilon \ll 1$, a Wentzel-Kramers-Brillouin (WKB) \cite{W,K,B} asymptotic expansion of the solution $\tau(x,c)\approx \tau(c)$  is known   \cite{KNESSL:1985bs,Matkowsky:1984,KNESSL:1984ij,Schuss:2010,Lagache:2012vn}
and can be written as 
\beq\label{taucc0}
\tau(c) \approx \frac{1}{r\left(x_0(c),c\right)}\frac{\ds{\sqrt{\frac{2\pi}{\epsilon
\frac{d}{dx}\left(\frac{l}{r}\right)\left(x_0(c),c\right)}}}}{\phi(x_T,c)},
\eeq
where $x_0(c)$ is the mean number of HA1 sites that are additionally protonated for a concentration $c>10^{-7}mol.L^{-1}$,  and the dependency of $x_0$ with respect to the concentration c in the range $0<x_0(c)=\epsilon X_0(c)<x_T$ has been obtained by a fitting procedure (see equation \ref{X0})
\beq
x_0(c)=\frac{7}{3}+\frac{\log(c)}{3\log(10)} \label{x0}. 
\eeq
Finally, by definition,
\beq
\phi(x,c)=\frac{\ds{\exp\left(-\frac{1}{\epsilon}\int_{x_0(c)}^{x}\log\left(\frac{l(s,c)}{r(s,c)}\right)ds \right)
}}{\ds\sqrt{\frac{l(x,c)}{r(x,c)}}} \left(\frac{l(x,c)}{r(x,c)}-1\right).\label{PHI}
\eeq
Now replacing the transition rates $r(x,c)$ and $l(x)$ by their expressions \ref{rl} in equation \ref{PHI} allows us to compute the mean  first passage time to the threshold. Indeed, 
\beqq
\int_{x_0(c)}^{x_T}\log\left(\frac{l(s,c)}{r(s,c)}\right) ds&=&\ds\int_{x_0(c)}^{x_T}\left(\log\left(10^{3s-7}\right)-\log(c)\right) ds\nonumber\\&=&\int_{x_0(c)}^{x_T}\left(3\log(10)s-\left(7\log(10)+\log(c)\right)\right) ds\label{log0},
\eeqq
that is
\beq
\int_{x_0(c)}^{x_T}\log\left(\frac{l(s,c)}{r(s,c)}\right) ds=\int_{x_0(c)}^{x_T}\left(3\log(10)s-\log(10^7c)\right) ds=F\left(x_T\right)-F\left(x_0(c)\right) \label{log},
\eeq
where
\beq
F(x)= \frac{3}{2}  \log(10) x^2 -\log(10^7c) x.\label{F}
\eeq
leading to
\beq
\phi(x_T,c)=\ds{\exp\left(-\frac{1}{\epsilon}\left(F\left(x_T\right)-F\left(x_0(c)\right)\right)\right)} \frac{\ds{
\frac{10^{3x_T-7}}{c}-1}}{\ds{{\sqrt{\frac{10^{3x_T-7}}{c}}}}}, \label{phi0}
\eeq
that is,
\beq
\phi(x_T,c)=\ds{\exp\left(-\frac{1}{\epsilon}\left(F\left(x_T\right)-F\left(x_0(c)\right)\right)\right)}  \ds{\left(
\frac{10^{3x_T/2-7/2}}{\sqrt{c}}-\sqrt{c} 10^{7/2-3x_T/2}\right)}.\label{phi}
\eeq
Using the expressions for the binding and unbinding rates \ref{rl}, we get
\beq
\frac{d}{dx}\left(\frac{l}{r}\right)\left(x_0(c),c\right)=\frac{d}{dx}\left(\frac{10^{3x-7}}{c}\right)\left(x_0(c),c\right)=\frac{3\log(10)}{c} 10^{3x_0(c)-7},
\eeq
which reduces to with formula \ref{x0} to
\beq
\frac{d}{dx}\left(\frac{l}{r}\right)\left(x_0(c),c\right)=3\log(10).\label{der}
\eeq
Finally, with eqs \ref{x0}-\ref{phi} and \ref{der}, we obtain for the mean conformational change time 
\beq
\tau(c) = \frac{\epsilon}{-Kc\left(\frac{4}{3}+\frac{\log(c)}{3\log(10)}\right)} \frac{\sqrt{\frac{\ds{2\pi}}{\ds{\epsilon
3\log(10)}}}\exp\ds{\left(\frac{1}{\epsilon}\left(F\left(x_T\right)-F\left(7/3+\frac{\log(c)}{3\log(10)}\right)\right)\right)}}{\ds{ \frac{10^{3x_T/2-7/2}}{\sqrt{c}}-\sqrt{c} 10^{7/2-3x_T/2}}}.
\eeq
Using that $\epsilon=1/n_s$ and $x_T=n_T/n_s$, we finally get the new expression for the conformational change time: 
\beq
\tau(c) = \frac{\sqrt{6\pi} \exp \left(\ds{n_s\left(F\left(n_T/n_s\right)-F\left(7/3+\frac{\log(c)}{3\log(10)}\right)\right)}\right)}{K \sqrt{cn_s\log(10)}\left(4+\frac{\log(c)}{\log(10)}\right) \ds{\left(c10^{{7/2-3n_T/(2n_s)}}-10^{{3n_T/(2n_s)-7/2}}\right)}},\label{taucc}
\eeq
where $F$ is defined in \ref{F}. 

Equation \ref{taucc} links the affinities between the ligand (concentration $c$) and the binding sites of a trimer to the conformational change mean time $\tau(c)$ of the trimer.  Interestingly, the reciprocal $\ds{\frac{1}{\tau(c)}}$ has been measured for various pH values \cite{Krumbiegel:1994bh}: $\left(\tau(pH=4.9)\right)^{-1}=5.78 s^{-1}$, $\left(\tau(pH=5.1)\right)^{-1}=0.12s^{-1}$,\ldots, $\left(\tau(pH=5.6)\right)^{-1}=0.017s^{-1}$.  Using formula \ref{taucc} and a least square optimization procedure, we have approximated the data (Figure 2-B) and obtain that the critical threshold is 
\beq
n_T\approx 6
\eeq
and the forward rate
\beq
K \approx 7.5 \hspace{0.2cm}10^3 L.mol^{-1}s^{-1}, \label{K}
\eeq
These two estimations can also been seen as predictions of the present model. We plotted in Figure \ref{figure2}-B the theoretical rate change for HA-conformational $\ds{\frac{1}{\tau(c)}}$ with respect to the proton concentration $c$ and compared it with the experimental values of \cite{Krumbiegel:1994bh}. We found a very good agreement (Figure \ref{figure2}-B), validating our jump-Markov model where the cumulative binding of few protons to an activating threshold $n_T$ leads to HA conformational change.

\subsection{A high potential barrier of HA binding sites ensures HA stability at neutral pH}
We have seen in section \label{protonBIG} that during endosomal acidification, a huge number of protons enter the endosome (more than  $20\hbox{ *}10^6$} and bind mostly to endosome buffers, leaving very few free protons (around $300$ at pH 6)). To test whether HAs buffer entering protons or interact with the remaining few free protons, we estimate the potential barrier generated at each HA binding site. For this purpose, we compare the reciprocal of the forward rate constant $K$ (equation \ref{K}), which is the mean time for a proton to bind a HA protein, with the free Brownian diffusion time scale. For a fixed proton concentration at a value $c$, the proton binding time is $\tau_{bind}=\frac{1}{Kc}$, while the mean time for a proton to diffuse to the same binding site is \cite{WARD:1993uq,SSH3,Schuss:2007fk,bookdavid}
\beq
\tau_{diff} \approx \frac{V}{4\pi D_{p} \eta} n(c)
\eeq 
. The number of endosomal protons at concentration $c$ is $n(c)=N_A c V $, while $\eta$ is the interacting radius between a proton and a binding site and $D_{p}$ the diffusion constant of a free proton ($D_p\approx 100 \mu m^2 s^{-1}$ measured in the cytoplasm \cite{al-Baldawi:1992ys}). For $\eta=1 nm$, we find a small ratio
\begin{linenomath}
\beq
\ds{\frac{\tau_{diff}}{\tau_{bind}}}=\frac{K}{4\pi D_{p} \eta N_A}\approx 10^{-4}. \label{NET}
\eeq
\end{linenomath}
confirming that the mean time for a proton to bind HA is dominated by a very high activation energy barrier at the HA binding sites, preventing rapid proton binding. Consequently, the buffering capacity of HAs can be neglected compared to the high capacity of other endosomal buffers. In addition, our model predicts that high HA1 potential barrier guarantees that the conformational change is only triggered after a cumulative binding of $n_T=6$ protons, ensuring a high stability of the protein at pH above 6, as previously characterized in Table 2 of \cite{Krumbiegel:1994bh} and confirmed in Figure S1.

In summary, we found that the threshold for HA1 conformational change occurs when there are $n_T=6$ bond proton in a total of $n_s=9$ binding sites. The binding is characterized by a very high potential barrier. Thus, when protons enter an endosome, they will first be captured by endosomal buffers. The remaining pool of free protons can bind to HA1 sites when they succeed passing the high potential barrier to ultimately trigger HA conformational change.

\section{A complete model of virus-endosome fusion}
Combining the kinetics model of endosome acidification with the Markov jump model of HA conformational change, we now propose a kinetics model of HAs conformational change inside an endosome. We account for the $n_T=6$ protons activating a HA1 trigger leading to HA conformational change.  We shall estimate the numbers $HA_0(t), HA_1(t) \ldots HA_6(t)$ of viral HAs that have $0,1 \ldots 6$ bound protons at time $t$, and compute the number of fusogenic (active) $HA_6(t)$, responsible for membrane fusion. From relation \ref{eqr}, the forward rate of a proton to a free HA1 binding site is
\begin{linenomath}
\beq
\tilde{r}\left(X\right)=r\left(X,P_e(t)\right)/P_e(t)=\frac{K n_s(n_s-X)}{N_A V_e} \label{rtilde}.
\eeq
\end{linenomath}
and the backward rate $l(X)$ is given by relation \ref{rl}, thus the chemical equations for protons $P_e$ and HA proteins are summarized by
\begin{linenomath}
\begin{equation*}
HA_0+P_e \overset{\tilde{r}(0/n_s)}{\underset {l(1/n_s)}{\rightleftharpoons}}  HA_1,
\end{equation*}
\begin{equation*}
HA_1+P_e \overset{\tilde{r}(1/n_s)}{\underset {l(2/n_s)}{\rightleftharpoons}}  HA_2,
\end{equation*}
\ldots \\
\begin{equation}
HA_5+P_e \xrightarrow{\tilde{r}(5/n_s)} HA_6. \label{chemical}
\end{equation}
\end{linenomath}
where the rate constant depends on each stage as given by relation \ref{rtilde}. The stage $HA_6$ is irreversible and the kinetic rate equations are
\begin{linenomath}
\begin{equation*}
\frac{dHA_0(t)}{dt}=-\tilde{r}\left(\frac{0}{n_s}\right)P_e(t)HA_{0}(t)+ l\left(\frac{1}{n_s}\right)HA_{1}(t),
\end{equation*}
\beq
\frac{dHA_1(t)}{dt}=\left(\tilde{r}\left(\frac{0}{n_s}\right)HA_0(t)-\tilde{r}\left(\frac{1}{n_s}\right)HA_{1}(t)\right)P_e(t) + l\left(\frac{2}{n_s}\right)HA_{2}(t)-l\left(\frac{1}{n_s}\right)HA_{1}(t),
\eeq
\ldots \\
\beq
\frac{dHA_6(t)}{dt}=\tilde{r}\left(\frac{5}{n_s}\right)HA_5(t) P_e(t). \label{differential}
\eeq
\end{linenomath}
Given the proton entry rate (equation \ref{acidif2}), these equations can be solved numerically.
\subsection{Modeling the onset of fusion between virus and endosome membranes}
The onset of membrane fusion is triggered by the conformational change of multiple adjacent trimer in the contact zone between virus and endosome membranes  \cite{Danieli:1996ve,Ivanovic:2013fk}. However, the number of fusogenic HAs involved in formation and fusion pore enlargement is still an open question. 

We model the contact zone between the virus and endosome membranes by $120$ HAs among the $400$ covering the virus  \cite{Ivanovic:2013fk}(Figure\ref{figure3}-A). Then, using a numerical solution of equation \ref{differential}, we chose randomly each new fusogenic HA and defines the onset of virus endosome fusion by the stochastic activation of $Na$ adjacent HAs in the contact zone (Figure \ref{figure3}-A). Using $1,000$ Monte-Carlo simulations, we estimated the mean and confidence interval at $95\%$ of the fusion onset time for different $Na$. 

We found that for $Na=1$ or $2$, most viruses fuse in EE, whereas for $Na=3$ or $4$ viruses fuse in ME. Finally, for $Na=5$ or $6$, viruses mostly fuse in LE (Figure \ref{figure3}-B,C). The common prediction is that $Na = 3- 4$ \cite{Danieli:1996ve,Ivanovic:2013fk} indicating that viruses shall fuse in ME. 
\subsection{Probing the intracellular localization of fusion with live cell imaging}
To determine the localization of virus fusion, we used  the fluorescent endosomal markers Rab5 (EE) and Rab7 (LE) in combination with an intracellular fusion assay to detect virus-endosome fusion so that the localization to a specific compartment can be assigned. Single virus spots were analyzed, where fusion was indicated by a pronounced increase of spot signal (Figure S2). To determine the cellular localization of virus fusion, we analyse infected Rab5- and Rab7-expressing cells with R18-labeled viruses (Figure \ref{figure3}-D). We classified single endosomes based on the presence of the two Rab proteins into three classes (Figure S3). Early endosomes (EE) do not show Rab7 association, such as late endosomes (LE) do not posses Rab5 signal. If endosomes possess both signals, they were counted as maturing endosomes (ME). We observe a gradual increase of Rab7 along with a decrease of Rab5 (Figure \ref{figure3}-D). After 5 min, we rarely observe fusion events in Rab5-only endosomes. The majority of fusion events (61\%) are detected in maturing endosomes between 10-20 min post infection (Figure \ref{figure3}-E). At later time points, the localization of fusion events shifted towards late endosomes. However, de-quenching kinetics show that fusion mostly occurs between 10-20 min (Figure S2). \\
We thus conclude that virus fusion was essentially associated with maturing endosomes indicating that $Na = 3 $ or $4$ adjacent fusogenic HA are needed to mediate fusion .

\section{Discussion}
Influenza  viruses are internalized into endosomes via receptor-mediated endocytosis. During their transport along microtubules, the endosomes accumulate protons, which eventually  enable virus-endosome fusion mediated by the influenza HA, resulting in release of the viral genome in the cell cytoplasm. Hence, the duration of endosomal transport as well as the  localization of fusion critically depend on endosomal acidification and HA conformational change at low pH. Here we presented a a new model to investigate the role of key parameters that shape the endosomal residence time of influenza viruses.

By associating a kinetics model of endosomal acidification with a Markov-jump process model of HA conformational change, we estimated the number of fusogenic HAs as function of time inside endosomes, and we modeled the onset of fusion with the stochastic activation of $Na$ adjacent HAs. Using the model, we predict the high HA stability at neutral pH due to the high activation barrier of protons binding sites. In association with $Na\geq 3$, this ensures that fusion occurs in ME, preventing a premature fusion in EE. As endosomal maturation is associated with retrograde transport of endosomes along MTs, this should increase the nuclear targeting of viral genome and pathogenicity of the virus.

\newpage


\begin{acknowledgements}
This research was supported by an Marie Curie grant (D.H.), by the Deutsche Forschungsgemeinschaft (HE 3763/15-1) (A.H.) and the Bundesministerium fur Bildung und Forschung  (eBio: ViroSign) (CS and AH). T.L. is funded by a Bourse Roux from Institut Pasteur.
\end{acknowledgements}

\bibliographystyle{spmpsci}
\bibliography{Reference_p1}

\pagebreak

\section*{Figures and tables}

\begin{figure}[!http]
\begin{center}
\includegraphics[angle=0,width=1.0\textwidth]{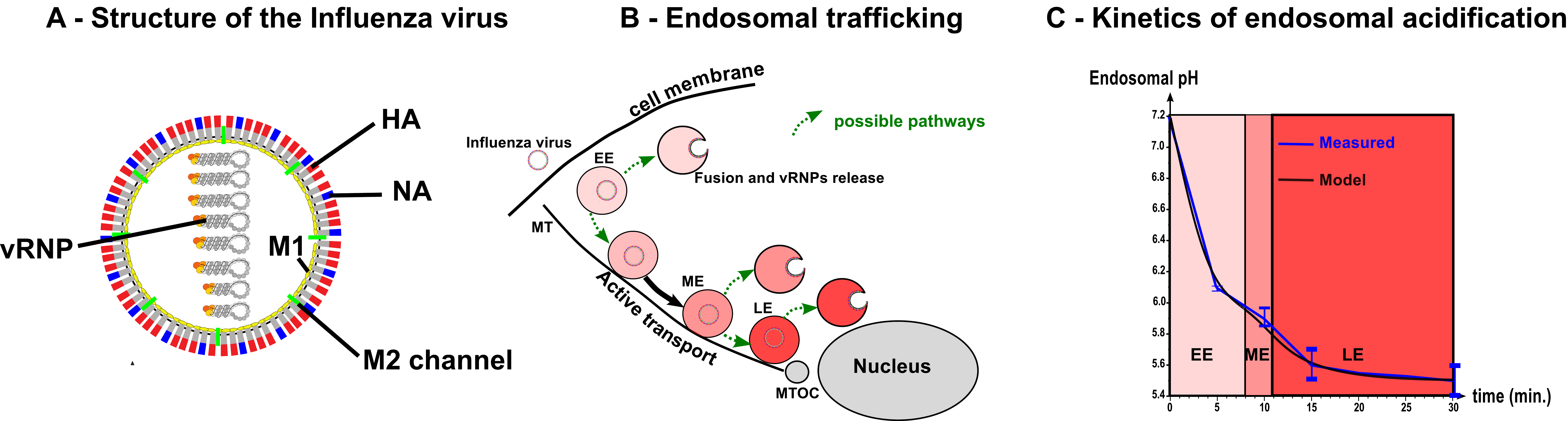}
\end{center}
\caption{\textbf{Structure and endosomal trafficking of the Influenza virus}{\bf A - Structure of the Influenza virus.} Influenza is an enveloped virus. Main spike proteins anchored in the envelope are the neuraminidase (NA) and the HA (HA). Protons can access the core of the virus through M2 channels. Main matrix protein is M1 protein. Viral genome of the virus is composed by eight viral ribonucleoproteins (vRNPs). {\bf B Endosomal trafficking of the virus.} Influenza virus enters the cell via receptor-mediated endocytosis and progress rapidly towards an Early Endosome (EE). Then, maturation of EE into a Maturing Endosome (ME) and Late Endosome (LE) is associated with an acidification of the endosome lumen and a retrograde transport of the endosome along the microtubules (MTs) of the cell towards the nucleus, the destination of vRNPS for virus replication. Fusion between the virus and the endosome membrane is critically controlled by the low pH conformational change of HAs, but the kinetics of in vivo escape remains largely unknown. {\bf C- Kinetics of endosomal pH decrease} obtained from intracellular fluorescence microscopy (red line. Mean $\pm$ SEM) and coarse-grained modeling (equation \ref{acidif2}, black line. Model parameters are summarized in table \ref{acidif_param}).}
\label{figure1}
\end{figure}

\begin{figure}[!http]
\begin{center}
\includegraphics[angle=0,width=1.0\textwidth]{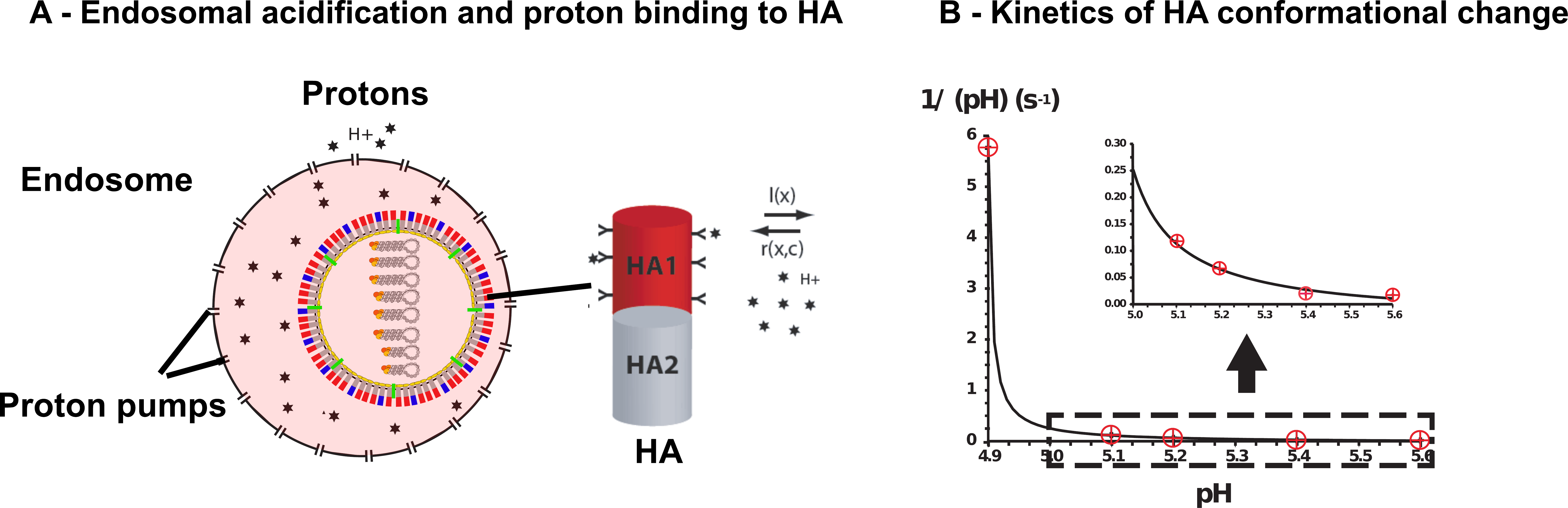}
\end{center}
\caption{{\bf Free protons in the endosome triggers HA conformational change} {\bf A -  Schematic representation of the influenza virus inside an endosome.} The right-hand side shows a scheme of an isolated HA trimer. Free protons in the endosome can bind to HA trimers. The protons binding rates $r(X,c)$ and $l(X)$ depend on the number of occupied sites $X$ and on the concentration $c$ of free protons in the endosome. When the number of bound protons reaches a given threshold, the HA trimer changes conformation into a fusogenic state. {\bf B Rate of the HA conformational change as a function of the pH.} The theoretical curve (solid line) for the rate of HA conformational change $\left(\tau(c)\right)^{-1}$ approximate well the experimental data (red circled crosses) \cite{Krumbiegel:1994bh}. The region inside the dashed box is magnified in the upper inset.}
\label{figure2}
\end{figure}

\begin{figure}[!http]
\begin{center}
\includegraphics[angle=0,width=1.0\textwidth]{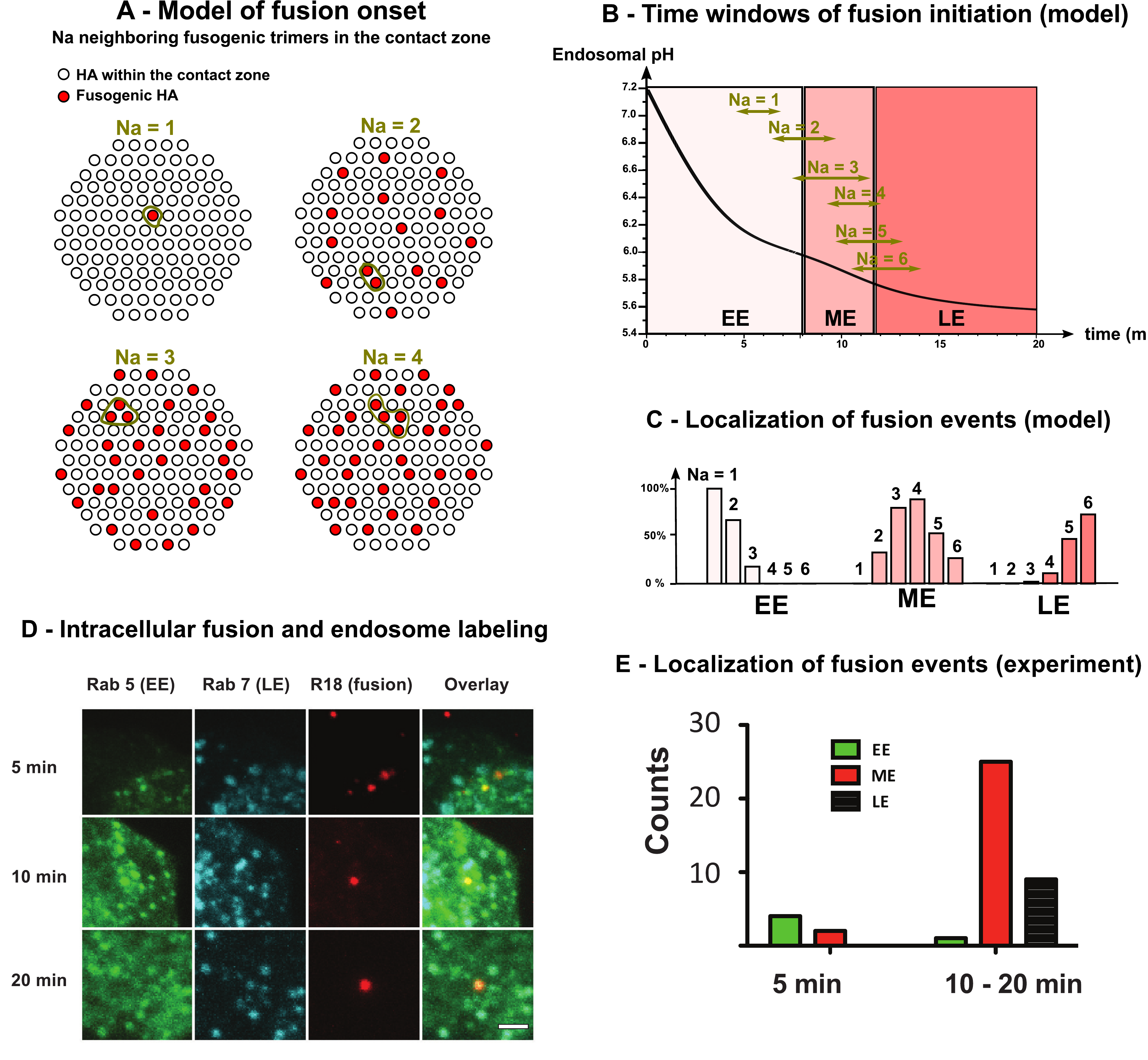}
\end{center}
\caption{{\bf Model and fluorescence experiments of the intracellular onset of virus-endosome fusion} {\bf A - Model of fusion onset.} The fusion between virus and endosome membranes is triggered by the conformational change of $Na$ adjacent HAs in the contact zone between virus and endosome ($\approx 120$ among the $400$ HAs covering the virus envelope \cite{Ivanovic:2013fk}). {\bf B - Modeling the stochastic activation of HAs on virus envelope during endosomal trafficking.} Solving equation \ref{differential} we estimated the time window ($95\%$ confidence interval) of intracellular fusion for $1\leq Na \leq 6$. {\bf C - Localization of fusion events as function of Na.} Using time windows of fusion onset and endosome maturation kinetics (equation \ref{RabC}), we estimated the localization (EE, ME or LE) of fusion onset as function of the number $Na$ of adjacent fusogenic HAs needed for the fusion onset.{\bf D - In vivo monitoring of fusion between virus and endosomes.} MDCK cells expressing  Rab5-CFP and Rab7-GFP were incubated with R18-labeled influenza A viruses. Fusion was observed as a strong increase of R18 signal due to de-quenching after dilution. Scale bar = 1 $\mu m$ {\bf E - In vivo localization of fusion events.} Fusion events were counted and categorized regarding their localization in EE (Rab5), ME (Rab5 + Rab7) or LE (Rab7).}
\label{figure3}
\end{figure}

\pagebreak

\begin{table}[http!]
\setlength{\extrarowheight}{3 pt}
   \caption{\bf{Parameters of the endosome acidification model}}
    \label{acidif_param}
\begin{tabular}{|l|m{7cm}|l|}
           \hline
    \textbf{Parameters} &\textbf{Description}& \textbf{Value}\tabularnewline
    \hline
     $r_e$ & Radius of the endosome & $r_e= 500\text{nm}$ \cite{Rink:2005kl} \tabularnewline
    \hline
         $V_e$ & Volume of the endosome & $V_e=\frac{4}{3}\pi r_e^3 = 5.22\hbox{ }10^{-16}\text{L}$ \tabularnewline
    \hline

         $r_v$ & Radius of the influenza virus & $r_v= 60\text{nm}$ \cite{Lamb:1983fk} \tabularnewline
          \hline
                   $V_v$ & Volume of the viral internal lumen & $V_v=\frac{4}{3}\pi r_v^3 = 9\hbox{ }10^{-19}\text{L}$ \tabularnewline
    \hline
                      $N_A$ & Avogadro constant & $N_A= 6.02\hbox{ }10^{23}\text{mol}^{-1}$ \tabularnewline
    \hline

      $\beta_e^0$ & Buffering capacity of the endosomal lumen & $\beta_e^0=40 \text{mM/pH}$ \cite{Van-Dyke:1994bh}  \tabularnewline
          \hline
      $\beta_v^0$ & Buffering capacity of the viral lumen & $\beta_v^0=\beta_e^0=40 \text{mM/pH}$ (this study) \tabularnewline
          \hline
          $\beta_v^{M1}$ & Buffering capacity of viral M1s & $\beta_{v}^{M1}\approx \frac{10500}{N_A V_v} \text{mM/pH}$ (this study) \tabularnewline
          \hline
         $\beta_v^{NP}$ & Buffering capacity of viral NPs & $\beta_{v}^{NP}\approx \frac{3000}{N_A V_v} \text{mM/pH}$ (this study) \tabularnewline
          \hline
         $\beta_v^{RNA}$ & Buffering capacity of viral RNA & $\beta_{v}^{RNA}\approx \frac{1200}{N_A V_v} \text{mM/pH}$ (figure 3-D in \cite{stoyanov}) \tabularnewline
          \hline
      $L^{\text{early}}$ & Permeability constant of early endosomes & $L^{\text{early}}\approx  3.5 \hbox{ } 10^{-3} N_A \text{cm s}^{-1}$ (this study) \tabularnewline
          \hline
      $L^{\text{late}}$ & Permeability constant of late endosomes & $L^{\text{late}}\approx  3.5 \hbox{ } 10^{-4} N_A \text{cm s}^{-1}$ (this study) \tabularnewline
      \hline
      $\text{pH}_{\infty}^{\text{early}}$ & Steady state pH of early endosomes & $\text{pH}_{\infty}^{\text{early}}=6.0$ \cite{Bayer:1998cr}  \tabularnewline
          \hline
      $\text{pH}_{\infty}^{\text{late}}$ & Steady state pH of late endosomes & $\text{pH}_{\infty}^{\text{late}}=5.5$ \cite{Bayer:1998cr} \tabularnewline
\hline
      $t_{1/2}$ & Half maturation time of endosomes & $t_{1/2}=10 \text{min.}$ (this study) \tabularnewline
\hline
           $\tau_{c}$ & Rab5/Rab7 mean conversion time & $\tau_{c}=100 \text{s}$ (figure 4-C in \cite{Rink:2005kl})  \tabularnewline
\hline
 \end{tabular}
\end{table}

\begin{table}[http!]
\setlength{\extrarowheight}{3 pt}
   \caption{\bf{Parameters of the HA's change of conformation model}}
    \label{HA_param}
\begin{tabular}{|l|m{7cm}|l|}
        \hline
    \textbf{Parameters} & \textbf{Description}& \textbf{Value}\\
    \hline
      $r(x,c)$ & Binding rate & $r(x,c)=K c n_s (1-x)$ (this study)\\
          \hline
      $l(x)$ & Unbinding rate & $l(x)=K n_s (1-x)10^{-\left(3\left(1-x\right)+4\right)}$ \cite{Huang:2002dq}\\
          \hline
      $n_T$ & Critical threshold for the number of HA1 bound sites & $n_T = 6$ (this study) \\
          \hline
      $K$ & Binding rate of a proton to a free HA1 binding site & $K=7.5*10^3 \text{L.mol}^{-1}\text{s}^{-1}$ (this study) \\
          \hline
      $n_s=1/\epsilon$ & Number of HA1 binding sites & $n_s=9$ \cite{Huang:2002dq}\\
          \hline
      $n_{HA}$ & Number of HAs &$n_{HA}= 400$  \cite{Ivanovic:2013fk}  \\
\hline
      \end{tabular}
 \end{table}

\end{document}